\documentclass[%
 reprint,
 amsmath,amssymb,
 aps,
superscriptaddress
]{revtex4-2}

\usepackage{graphicx}
\usepackage{dcolumn}
\usepackage{bm}
\usepackage{gensymb}

\usepackage{textcomp}
\usepackage{xspace}
\usepackage{notes2bib}

\newcommand{\murm}{%
  \ifmmode
    \mathchoice
        {\hbox{\normalsize\textmu}}
        {\hbox{\normalsize\textmu}}
        {\hbox{\scriptsize\textmu}}
        {\hbox{\tiny\textmu}}%
  \else
    \textmu
  \fi
}

\newcommand{\micrometer}{\murm{\rm m}\xspace}
\newcommand{\microsecond}{\murm{\rm s}\xspace}

\begin{document}

\preprint{APS/123-QED}

\title{3D-Printed Micro Ion Trap Technology for Scalable Quantum Information Processing}

\author{Shuqi Xu}
\email{These two authors contributed equally.}
\affiliation{Department of Physics, University of California, Berkeley, Berkeley, CA 94720, USA}
\affiliation{Challenge Institute for Quantum Computation, University of California, Berkeley, Berkeley, CA 94720, USA}

\author{Xiaoxing Xia}
\email{These two authors contributed equally.}
\affiliation{Engineering Directorate, Lawrence Livermore National Laboratory, Livermore, CA 94550, USA}

\author{Qian Yu}
\affiliation{Department of Physics, University of California, Berkeley, Berkeley, CA 94720, USA}
\affiliation{Challenge Institute for Quantum Computation, University of California, Berkeley, Berkeley, CA 94720, USA}

\author{Sumanta Khan}
\affiliation{Department of Physics, University of California, Berkeley, Berkeley, CA 94720, USA}
\affiliation{Challenge Institute for Quantum Computation, University of California, Berkeley, Berkeley, CA 94720, USA}

\author{Eli Megidish}
\email{Current address: Atom Computing, Inc., Berkeley, CA 94710, USA.}
\affiliation{Department of Physics, University of California, Berkeley, Berkeley, CA 94720, USA}
\affiliation{Challenge Institute for Quantum Computation, University of California, Berkeley, Berkeley, CA 94720, USA}

\author{Bingran You}
\affiliation{Department of Physics, University of California, Berkeley, Berkeley, CA 94720, USA}
\affiliation{Challenge Institute for Quantum Computation, University of California, Berkeley, Berkeley, CA 94720, USA}

\author{Boerge Hemmerling}
\affiliation{Department of Physics and Astronomy, University of California, Riverside, Riverside, CA 92521, USA}

\author{Andrew Jayich}
\affiliation{Department of Physics, University of California, Santa Barbara, Santa Barbara, CA 93106, USA}

\author{Juergen Biener}
\email{biener2@llnl.gov}
\affiliation{Physical and Life Science Directorate, Lawrence Livermore National Laboratory, Livermore, CA 94550, USA}

\author{Hartmut~H\"affner}%
\email{hhaeffner@berkeley.edu}
\affiliation{Department of Physics, University of California, Berkeley, Berkeley, CA 94720, USA}
\affiliation{Challenge Institute for Quantum Computation, University of California, Berkeley, Berkeley, CA 94720, USA}
\affiliation{Computational Research Division, Lawrence Berkeley National Laboratory, Berkeley, CA 94720, USA}

\begin{abstract}

Trapped-ion applications, such as in quantum information, precision measurements, optical clocks, and mass spectrometry, rely on specialized high-performance ion traps. The latter applications typically employ traditional machining to customize macroscopic 3D Paul traps, while quantum information processing experiments usually rely on photo-lithographic techniques to miniaturize the traps and meet scalability requirements. 
Using photolithography, however, it is challenging to fabricate the complex three-dimensional electrode structures required for optimal confinement. Here we address these limitations by adopting a high-resolution 3D printing technology based on two-photon polymerization supporting fabrication of large arrays of high-performance miniaturized 3D traps. We show that 3D-printed ion traps combine the advantages of traditionally machined 3D traps with the miniaturization provided by photolithography by confining single calcium ions in a small 3D-printed ion trap with radial trap frequencies ranging from $2$\,MHz to $24$\,MHz. The tight confinement eases ion cooling requirements and allows us to demonstrate high-fidelity coherent operations on an optical qubit after only Doppler cooling. With 3D printing technology, the design freedom is drastically expanded without sacrificing scalability and precision so that ion trap geometries can be optimized for higher performance and better functionality.

\end{abstract}

\maketitle

\section*{Main}

\begin{figure*}%
\centering
\includegraphics[width=1\textwidth]{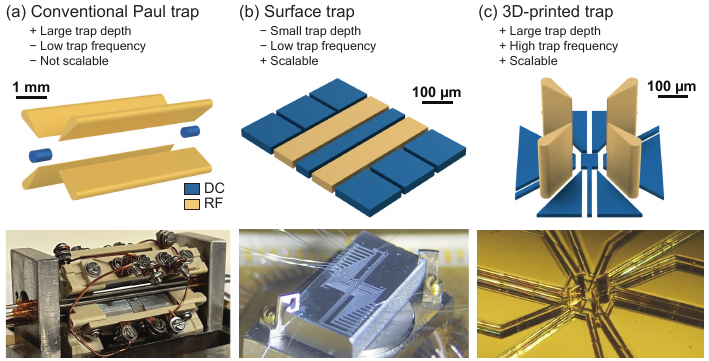}
\caption{\label{fig1}\textbf{Paul trap comparison.} Schematics and representative pictures of a conventional 3D 
Paul trap {\bf a}, a surface trap {\bf b}, and a 3D-printed vertical Paul trap {\bf c}. The basic structure of a 
conventional 3D Paul trap (\textbf{a}) consists of four RF electrodes and two end cap DC electrodes, with a characteristic length scale of 1\,mm. It requires precision machining and is thus challenging to scale. Surface traps (\textbf{b}) usually have two RF electrodes with a characteristic length scale on the order of 100\,\micrometer and can be produced by microfabrication techniques allowing for large and scalable trap arrays. 
However, constraining electrodes to a single plane distorts the desired quadrupole potential reducing its efficiency. 3D-printed Paul traps (\textbf{c}) can have four RF electrodes creating a quadrupole field while of similar size and scalability as surface traps. }
\end{figure*}

\begin{figure}%
\centering
\includegraphics[]{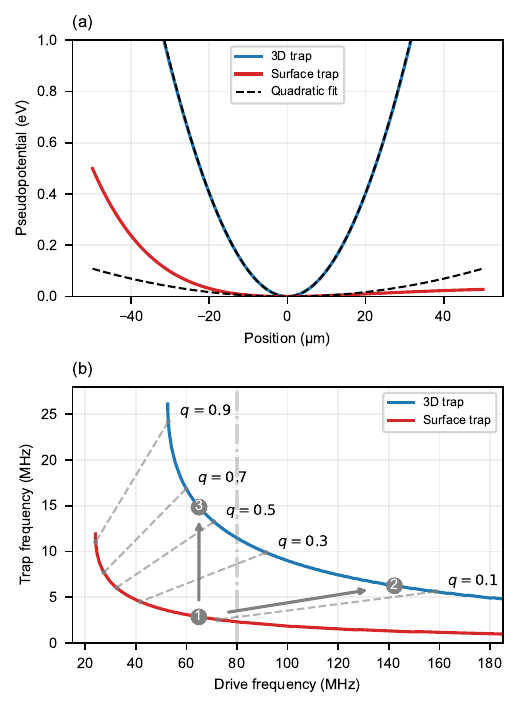}
\caption{\textbf{Performance comparison between 3D and surface traps}. We use the geometries sketched in Fig.\,\ref{fig1}a and b and set the ion-to-electrode distances to 100\,\micrometer. The two pairs of opposing RF electrodes of the 3D trap are driven with opposite polarity. \textbf{a}, Pseudopotential (see methods) for the 3D (blue) and surface trap (red) configurations with 80\,MHz drive frequency and 150\,V RF voltage amplitude. The dashed lines are harmonic fits. 
The 3D trap is much more harmonic than the surface trap and consequently has a much larger trap depth. The confinement is also tighter for the 3D trap as compared to the surface trap. \textbf{b}, Trap frequencies as a function of drive frequency (solid lines), assuming that the RF voltage amplitude $\widetilde{U}=150\,\rm{V}$ is constant. Dashed lines indicate a constant stability parameter $q$ (see Eq.\,\ref{eq:definition-q} in the Methods). The vertical dash-dotted line at $80\,$MHz indicates the parameter set used in Fig.\,\ref{fig2}\textbf{a}. Point 1 corresponds to a set of viable surface trap parameters. Point 2 (3) indicates parameters when holding the stability parameter $q$ (drive frequency) constant. For the 3D trap, we predict doubling of the trap frequency when holding $q$ constant and a five-fold increase when keeping the drive frequency constant.\label{fig2}}
\end{figure}

Ion traps are an important tool in a wide range of fields \cite{leibfried_quantum_2003} such as mass spectroscopy \cite{douglas_linear_2005}, precision metrology \cite{Brown1986a, Smorra2017-anti-proton,scielzo_-decay_2012, Roussy2023}, optical ion clocks \cite{ludlow_optical_2015}, and quantum information science \cite{Bruzewicz2019-ion-trap-review}. Until about a decade ago, charged particles have been typically confined in macroscopic ion traps \cite{paul_electromagnetic_1990}.
These macro 3D traps feature a near harmonic trapping potential, high trapping efficiency, and large trap depth that keeps the particles in a deep potential well (see Fig.\,\ref{fig1}a). However, machining constraints limit fabrication of miniaturized, complex electrode structures that are required for building large-scale trapped ion quantum processors. Furthermore, the relatively large ion-to-electrode distance of macro 3D traps ($\sim$1\,mm) also limits the electric field strength for a given voltage and thus trap frequency.

These difficulties motivate the development of surface traps based on planar electrode structures amenable to microfabrication \cite{Chiaverini2005, Seidelin2006} (see Fig.\,\ref{fig1}b). Surface traps are compatible with well-established micro-electromechanical systems (MEMS) and complementary metal-oxide semiconductor (CMOS) microfabrication techniques and thus allow for miniaturization and scaling to large arrays with complex 2D structures, while also enabling integrated photonic ion-light interfaces. Compared to 3D geometries, the arrangement of surface trap electrodes in a single plane leads to large anharmonicities in the potential and reduces the trap depth substantially \cite{Chiaverini2005, Wesenberg2008}. 
In addition, the deviation from the quadrupole potential for 2D geometries reduces the trap frequency, which in turn requires trapping the ions closer to the electrodes to maintain reasonable trap frequencies. The proximity of the ion to the electrode surfaces exposes the ion to electric field noise caused by the electrodes. This noise heats up the ion motion and can be a major source of errors for quantum gates \cite{brownnutt_ion-trap_2015,Brown2021-materials}. In order to overcome these challenges, there have been efforts to build 3D ion traps compatible with microfabrication techniques using stacked wafers \cite{Blakestad2009, ragg_segmented_2019, decaroli_design_2021}, but the design flexibility remains limited \cite{See2013-monolithic-traps,auchter_industrially_2022}.

Here we demonstrate a novel approach to fabricate miniaturized 3D Paul traps (Fig.\,\ref{fig1}c) that combines the efficiency of macro 3D traps with the scaling advantages of surface traps \cite{biener_miniature_2022,quinn_geometries_2022}. In particular, we use high-resolution 3D printing based on two-photon polymerization \cite{baldacchini_three-dimensional_2016} which enables complex, micro- and nano-architected designs with sub-micron resolution for a variety of applications \cite{xia_electrochemically_2019, gao_high-resolution_2020, oellers_-chip_2019, fendler_microscaffolds_2019}. 
Compared to surface traps, 3D-printed Paul traps achieve a larger trap depth, a more harmonic trapping potential and higher trap frequencies.

As the main performance criteria for 3D and surface traps, we use the depth and curvature of their confining effective potential. For a sufficiently large drive frequency of the rapidly oscillating trapping RF field, the confinement can be described by a static ponderomotive potential (pseudopotential) \cite{Leibfried2003review}.
Figure\,\ref{fig2}a compares the pseudopotential of a miniaturized 3D trap to a surface trap, keeping the ion-to-electrode distance, the drive frequency, and the amplitude of the oscillating trapping RF field constant. Especially visible is the dramatic reduction in the pseudopotential depth of surface traps. Further, we find that the pseudopotential of the 3D configuration is much more harmonic and provides a larger trap frequency than its planar counterpart.

To evaluate the expected performance gain, we must also account for the trapping mechanism for Paul traps.
In particular, the pseudopotential fails to describe the motional frequencies of the ion in the effective trap potential when the motional frequency approaches one fifth of the frequency of the applied trapping RF field. This breakdown is characterized by the so-called stability parameter, $q$ (see Eq.\,\ref{eq:definition-q} in the Methods), and occurs at approximately $q=0.5$. While micro 3D traps with their harmonic potential are expected to tolerate a larger stability parameter $q$ than surface traps, the higher motional frequency may lead to trap instabilities. 
To compensate for this, one may increase the drive frequency to maintain a constant $q$. 
In Fig.\,\ref{fig2}b, we illustrate the associated trade-offs assuming the same RF voltage amplitude $\widetilde{U}$, for both surface and 3D configurations. 
Under this assumption, we compare the calculated trap frequency holding either the drive frequency or $q$ (diagonal dashed lines) constant. We see that for the same $\widetilde{U}$, micro 3D traps offer indeed larger trap frequencies. In particular, assuming typical parameters for a surface trap indicated by Point 1, the same drive frequency will lead to a five-fold increase in trap frequency for the 3D trap (Point 3 vs. Point 1). If one keeps the stability parameter $q$ constant, the trap frequency is doubled for the micro 3D trap (Point 2 vs. Point 1). Increasing the trap frequency is beneficial in many aspects as it reduces motional heating, allows for faster ion movement, and reduces cooling complexity. Alternatively, one may prioritize power consumption, which is important for large arrays of traps. Here, the micro 3D trap configuration achieves the same trapping parameters (trap frequency and stability parameter) with about an order of magnitude less power than surface trap designs. Another option to utilize the higher efficiency of 3D traps is to increase the ion-to-electrode distance, $d$, to reduce surface noise while maintaining a similar trap frequency.

\section*{Results}
\begin{figure*}%
\centering
\includegraphics[width=0.8\textwidth]{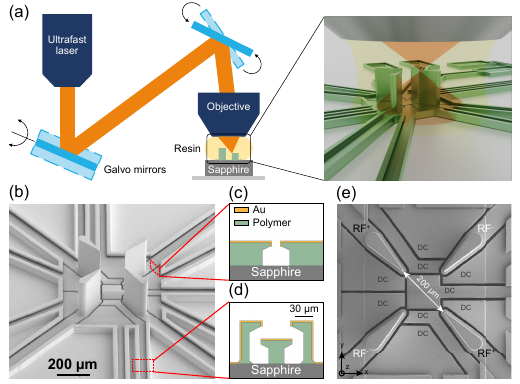}
\caption{\textbf{3D printing process and SEM images of fabricated traps.} \textbf{a}: Illustration of the 3D printing method: an objective focuses pulsed femtosecond laser light at 780\,nm (orange) into a liquid photoresist (yellow). Galvo mirrors are used to raster scan the focal point to print the designed trap geometry (green) which is subsequently coated with metal. \textbf{b}: the SEM image of a vertical linear Paul trap. \textbf{c} shows the cross section schematic of the shadow mask design between electrodes (highlighted via the top red dashed rectangle in \textbf{b}). \textbf{d} shows the cross section schematic of the electrical pathways including side walls to improve robustness with respect to electrical isolation. \textbf{e} shows an SEM image of a trap from the top. The two ``$\rm{RF^+}$'' electrodes carry the same RF signal as the two ``$\rm{RF^-}$'' electrodes but with opposite polarity. Static voltages are applied to the ``DC'' electrodes.}\label{fig3}
\end{figure*}

In this work, we design and fabricate a 3D-printed micro linear Paul trap using a commercial Nanoscribe two-photon lithography system (see Fig.\,\ref{fig3}a, Methods) and test its performance by confining ${ }^{40} \rm{Ca}^{+}$ ions at room temperature. 
The trap consists of four RF electrode pillars on a sapphire substrate with a total height of 300\,\micrometer (see Fig.\,\ref{fig3}b). The distance between opposing RF electrodes is 200\,\micrometer, resulting in a 100\,\micrometer ion-to-RF-electrode distance. As shown in Fig.\,\ref{fig3}e, nine planar DC electrodes are placed on the substrate surface for tuning the potential in a 600\,\micrometer $\times$ 600\,\micrometer square. We use a configuration where neighboring RF electrodes are driven out-of-phase, while opposing electrodes are in-phase with respect to each other. By choosing an identical amplitude for all RF electrodes their vertical electric fields cancel out. The nine DC electrodes provide confinement in the vertical direction, allow one to cancel electric stray fields, and the ability to choose the orientation of the effective quadrupole potential.

The trap is fabricated by first creating 3D polymer structures using two-photon polymerization and then coating the polymer with a 1\,\micrometer-thick gold layer using electron-beam evaporation (Fig.\,\ref{fig3}a, Methods). 
To isolate individual electrodes, we use an undercut below the electrode top surfaces to keep them electrically isolated after gold coating. The undercut cross section is shown in Fig.\,\ref{fig3}c. The electrical pathways connecting the DC and RF electrodes are 3D-printed using a similar shadowing method to allow for routing flexibility. To increase the robustness for electrical isolation, we add an overhanging sidewall on both sides of the T-shaped cross section of the electrical paths as shown in Fig.\,\ref{fig3}d. 
The flexibility of printing trap structures and electrical paths in this maskless one-step process results in a turnaround time from design to a working device of one to two days. These short turnaround times enable us to iterate and improve trap designs rapidly. 

\begin{figure}%
\centering
\includegraphics[]{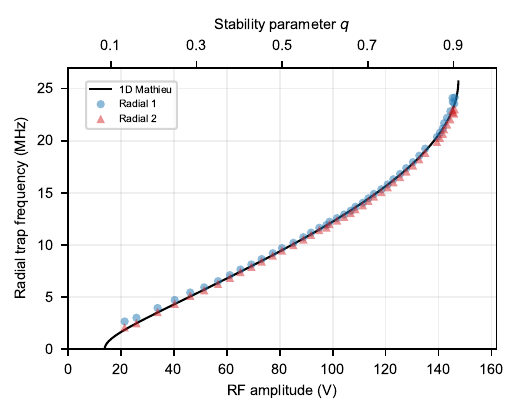}
\caption{\textbf{Radial trap frequencies.} 
Blue circles and red triangles show the two measured radial trap frequencies split by a DC quadrupole field corresponding to $\alpha=0.0018$ (see Methods) as a function of the RF amplitude measured with a capacitive divider up to a scale factor. The black line is a numerical solution of the Mathieu differential equation (Methods, Eq.\,\ref{eq:mathieu}) for the trap frequency as a function of the stability parameter $q$. We use the fact that the measured voltage at the capacitive divider is proportional to $q$ to determine the corresponding proportionality constant by fitting the numerical solution of the Mathieu equation to the data. This allows us then to plot the measured frequencies as a function of $q$. 
We measured trap frequencies for $q$ ranging from 0.13 to 0.9 and derived the actual applied RF amplitude from electrostatic simulations and the stability parameter $q$. }\label{fig4}
\end{figure}

For trapping, we drive the RF electrodes with frequency $\omega_{\rm{rf}}/2\pi = 51.6\,\rm{MHz}$.
Depending on the chosen RF voltage amplitude, we observe radial trap frequency $\omega/2\pi$ ranging from $2.09\,\rm{MHz}$ to $24.15\,\rm{MHz}$ (see Fig.\,\ref{fig4}). 
At 24.15\,MHz, our highest measured trap frequency, $q = 0.903$ (see Methods), near the theoretical limit of 0.911 ($a=0.0018$), demonstrating operation throughout the stability diagram. 

\begin{figure}%
\centering
\includegraphics[]{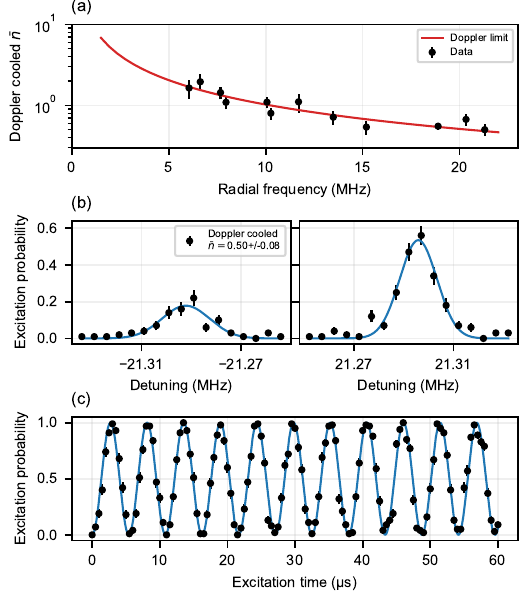}
\caption{\textbf{Cooling and gate operations.} 
The black dots in \textbf{a}, \textbf{b}, and \textbf{c} are data. \textbf{a}, the resulting occupation number, $\bar{n}$, at different radial trap frequencies. The red line is the theoretical Doppler cooling limit, taking into account the 45$^\circ$ overlap between the cooling laser $\vec{k}$-vector and the motional mode direction. Error bars are calculated from the $\chi^2$-fit of the amplitude. \textbf{b}, the harmonic oscillator occupation $\bar{n}$ after Doppler cooling is extracted via spectroscopy of the motional sidebands at $\omega/2\pi = \pm 21.29\,\rm{MHz}$, where the lines are Gaussian fits. \textbf{c}, single qubit Rabi oscillations performed at a radial trap frequency of $21.29\,\rm{MHz}$ with only Doppler cooling where the fit gives a Rabi frequency of 2$\pi\times$185\,kHz. Error bars in \textbf{b} and \textbf{c} indicate quantum projection noise. 
}\label{fig5}
\end{figure}

We further characterize the cooling performance at various trap frequencies. For this, we measure the motional occupation of one of the two planar modes oriented 45 degrees with respect to the horizontal main cooling beam near 397\,nm. The average harmonic oscillator quantum number $\bar{n}$ is determined via laser spectroscopy on the $S\leftrightarrow D$ transition. If the ion is near the motional ground state, absorption on the red sideband (the laser detuning $\Delta = -\omega/2\pi$) associated with annihilating a motional quantum is suppressed as compared to absorption on the blue sideband ($\Delta = + \omega/2\pi$) which leads to the creation of a motional quantum \cite{Leibfried2003review}. Figure\,\ref{fig5}a explores the temperature after Doppler cooling for different radial trap frequencies. The data matches well with the theoretical Doppler cooling limit with no free parameters \cite{leibfried_quantum_2003}. In particular, Fig.\,\ref{fig5}b shows the measured red and blue sideband spectra at $\Delta =\pm\omega/2\pi = \pm 21.29$\,MHz after Doppler cooling revealing that an average quantum number of $\bar{n}=0.5$ is reached.

For quantum control, small motional excitation is important as the motion modulates the laser frequency experienced by the ion thereby causing gate errors \cite{Wineland1998-experimental-issues,Sutherland2022-gate-errors}. For instance, the carrier Rabi frequency which determines the speed of single-qubit operations is \cite{wineland_experimental_1998}
\begin{equation}\label{eq:thermal-Rabi-flops}\Omega = \Omega_0\left(1-\sum_i n_i \eta_i^2\right)\,,
\end{equation} where $\eta_i$ is an effective Lamb-Dicke parameter, $n_i$ is the motional quantum number of mode $i$ for the addressed ion, and $\Omega_0$ is the overall coupling strength. In existing trapped ion quantum computers this can cause gate errors on the order of $10^{-3}$. For this reason often a number of motional modes are cooled to 
the ground state to reduce the error rate \cite{Schindler2013}. The large trap frequencies of 3D-printed traps mitigates this problem. In particular, assuming a trap frequency of 20\,MHz 
and the Ca$^+$ optical qubit used here near the Doppler limit, this effect is expected to add only $2\times10^{-7}$ to the error rate of $\pi$ pulses.
Figure\,\ref{fig5}c shows Rabi oscillation measured at $21.29$\,MHz trap frequency after only Doppler cooling. The contrast of the Rabi oscillations decay from $0.994^{+0.006}_{-0.010}$ for the first oscillation to $0.993^{+0.007}_{-0.031}$ for the 11$^{\rm th}$ oscillation 
implying error rates of $\lesssim 10^{-4}$ for a $\pi$-rotation.
The reduction in contrast can be explained by laser intensity noise and residual overlap between the uncooled (vertical) axial motion and the 729\,nm beam. 

\section*{Discussion}
While very high trap frequencies have been achieved with lighter ions \cite{jefferts_coaxial-resonator-driven_1995}, our trap frequencies exceed typical values, both for macro 3D traps \cite{schindler_quantum_2013} and surface traps by a factor of four. 
As a result, the operational timescales such as splitting, merging and shuttling of ion crystals would be accelerated in such traps.

Further, we expect that the impact of surface noise is reduced. In particular, motional heating is expected to scale as $1/\omega^{(1+\lambda)}$, with the frequency exponent $\lambda$ of electric field noise to be of order 1-1.5 \cite{Brown2021-materials}. Thus heating is expected to be substantially suppressed at higher trap frequencies.

In addition, higher trap frequencies also promise to reduce cooling requirements and thereby to accelerate the cooling process itself. 
In particular, both $\bar{n}$ as well as $\eta$ are reduced with increasing trap frequency such that according to Eq.\,\ref{eq:thermal-Rabi-flops} the impact of a finite motional occupation on the gate fidelity is expected to scale with $1/\omega^2$. 
We estimate that for the optical qubit in $^{40}$Ca$^+$ axial trap frequencies of 10\,MHz are sufficient to achieve error rates of below $10^{-5}$ solely with Doppler cooling. We believe that this regime is well within reach with miniaturized 3D traps. For instance, assuming an ion-to-RF-electrode distance of order 50\,\micrometer, a drive of amplitude 160\,V and frequency 150\,MHz ($q=0.5$), yields a radial trap frequency of 30\,MHz for Ca$^+$-ions, sufficient to maintain linear orientation for a few ions at axial frequencies of 10\,MHz.
Thus, future quantum computers would not require sideband cooling, which reduces the cooling cycles from many milliseconds to a few 100\,\microsecond \cite{leibfried_quantum_2003}.  
Since cooling, splitting, merging, and shuttling occupy most of the duty cycle of current trapped ion quantum computers based on a Quantum Charge-Coupled Device (QCCD) architecture \cite{Home2006,pino_demonstration_2021,moses_race_2023}, such computers could be sped up significantly using 3D-printed ion traps.

\begin{figure}%
\centering
\includegraphics[]{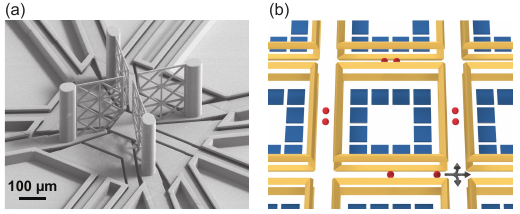}
\caption{\textbf{Outlook.} \textbf{a}, SEM image of a 3D-printed vertical linear Paul trap with meshed RF electrodes to reduce surface area and ion-to-RF-electrode distance. \textbf{b}, Schematic of 3D-printed QCCD architecture. RF electrodes are in yellow, while DC electrodes are in blue. It consists of multiple trapping zones connected via X-junctions. Each trapping zone is a micro horizontal linear Paul trap. Ions (red) can be shuttled between zones assisted by DC electrodes (blue) through X-junctions, as illustrated by gray arrows.}\label{fig6}
\end{figure}

The successful trapping of ions in a 3D-printed ion trap opens a new path towards flexible and scalable miniaturized ion traps. 
The novel design space allows one to explore trap designs that cannot be realized using traditional fabrication methods. In Fig.\,\ref{fig6}a, we show an example of a printed trap with significant reduction of the surface area near the ion. It would be interesting to study how such extreme geometries impact the electric surface noise experienced by the ion \cite{Low2011-distance-scaling} and thus to learn more about the causes for surface electric field noise. 

The driving motivation for our work is to enable high-performance trapped-ion quantum computers, especially based on the QCCD architecture (see Fig.\,\ref{fig6}b). The 3D-printed ion trap platform offers multiple options towards this goal. One may leverage the higher trap efficiency to 1. increase confinement which reduces cooling requirements and accelerates splitting and merging of ion crystals, 2. increase the ion-to-electrode distance and thus substantially reduce motional heating, or 3. reduce trap drive power to ease thermal management, which is especially relevant in future large-scale arrays. Finally, a weighted combination of 1., 2. and 3. might be beneficial for specific realizations. 
Thus, 3D-printed traps appear to be an attractive technology for large scale devices. In this context, we estimate that array densities in excess of 1,000 traps per cm$^2$ can be fabricated, ideally on substrates that house integrated photonic circuits to control the ions to allow for scalable optical control \cite{niffenegger_integrated_2020,mehta_integrated_2020}.

Miniaturized 3D-printed traps may also impact other disciplines than quantum information processing. For instance, we envision that they may be used as ultra-compact low-power mass spectrometers attractive for space applications. In precision metrology, 3D-printed trap arrays may improve the signal-to-noise ratio substantially and thus would increase the stability of, for instance, optical ion clocks \cite{ludlow_optical_2015}. More fundamentally, the superior harmonicity of 3D traps allows for stable trap frequencies without laser cooling which is also a very important consideration for precision metrology applications \cite{Brown1986a}. 
Trap anharmonicity has also been a major hurdle for past efforts for quantum information with trapped electrons in Penning traps \cite{Bushev2008,Goldman2010}. Here, the efficiency of harmonic 3D traps provides a critical component for current efforts to use trapped electrons for fast and high-fidelity quantum information processing \cite{yu_feasibility_2022}.

In summary, we have designed and fabricated 3D-printed ion traps that combine high trap frequency and deep harmonic potentials with miniaturization and scalability. With continuous improvement in resolution and speed, 3D printing opens
a new vista of ion trap development with dramatically expanded geometric freedom and submicron control of features that may be
optimized for functionality beyond the limits of photolithography. We envision that the 3D-printed ion trap platform with integrated photonics would accelerate the development of quantum information processing systems as well as on-chip mass spectroscopy, precision metrology, and optical clocks, especially if 3D printing can be integrated into foundry-based microfabrication workflow.

\begin{acknowledgments}
The authors acknowledge Clemens Matthiesen's inspiration at the start of the project and James Oakdale for printing the first demonstration objects. The authors thank Kristin Beck for helpful discussions. The authors would also like to acknowledge Nicole Greene for the help on chamber assembly, Wei-Ting Chen for the useful guidance on the measurements, Ben Saarel for providing the image in Fig.\,\ref{fig1}b, and Abhinav Parakh for rendering the illustration in Fig.\,\ref{fig3}a. This work is supported by the UC Multicampus-National Lab Collaborative Research and Training under Award No.~LFR-20-653698.
Part of this work was performed under the auspices of the U.S. Department of Energy by Lawrence Livermore National Laboratory under Contract DE-AC52-07NA27344. AJ acknowledges the support of ONR Grant No. N00014-21-1-2597.
\end{acknowledgments}

\bibliography{references, sn-bibliography}

\clearpage
\newpage
\section*{Methods}
\subsection*{Paul trap equations}
In Paul traps, ions experience a static potential plus a time-dependent potential \cite{leibfried_quantum_2003}:

\begin{equation}
    \begin{aligned} \Phi(x, y, z, t) & = \Phi_{\rm{static}} + \Phi_{\rm{rf}} \\ & =\frac{U}{2}\left(A x^{2}+B y^{2}+C z^{2}\right) \\ & + \frac{\widetilde{U} \cos \left(\omega_{\mathrm{rf}} t\right)}{2}\left(A^{\prime} x^{2}+B^{\prime} y^{2}+C^{\prime} z^{2}\right)\ .
    \end{aligned}
\end{equation}
The coefficients $A, B, C$ ($A^{\prime}, B^{\prime}, C^{\prime}$) satisfy the Laplace equation $\Delta \Phi_{\rm{static}} = 0$ ($\Delta \Phi_{\rm{rf}} = 0$).
The motion of a particle with mass $m$ and charge $Ze$ in the $x$-direction is described by the following equation:

\begin{equation}
    \begin{aligned}
        \ddot{x}=-\frac{Ze}{m} \frac{\partial \Phi}{\partial x}=-\frac{Ze}{m}\left[U A+\widetilde{U} \cos \left(\omega_{\mathrm{rf}} t\right) A^{\prime}\right] x\ .
    \end{aligned}
\end{equation}
Further simplification leads to the standard Mathieu differential equation:

\begin{equation}\label{eq:mathieu}
    \frac{d^{2} x}{d \xi^{2}}+\left[a-2 q \cos (2 \xi)\right] x=0
\end{equation}
with $\xi=\omega_{\mathrm{rf}} t / 2, a=4 Z|e| U A / m \omega_{\mathrm{rf}}^{2}$ and
\begin{equation}\label{eq:definition-q}
q=-2 Z|e| \widetilde{U} A^{\prime} / m \omega_{\mathrm{rf}}^{2}\,.    
\end{equation}
Without loss of generality, we can set $q\geq 0$. The stable solution of the Mathieu equation has a characteristic exponent $\beta$ which depends on $a$ and $q$ leading to trap frequency solution $\nu=\beta\xi/2\pi$. In the lowest-order approximation where $\left(\lvert a \rvert, q^{2}\right) \ll 1$,
\begin{equation}
    \beta \approx \sqrt{a+q^{2} / 2}\ .
\end{equation}
The ion's motion can be described by a ponderomotive potential, which is also known as the pseudopotential:
\begin{equation}
    U_{\mathrm{ps}}=\frac{1}{4 m \omega_{\mathrm{rf}}^{2}}\left(\mathbf{F}_{\mathbf{r f}}\right)^{2}\ .
\end{equation}
$\mathbf{F}_{\mathbf{r f}}$ is the magnitude of the RF force acting on the ion. When the lowest-order approximation is no longer valid, i.e.~when $q\approx 0.5$, the characteristic exponent $\beta$ can be extracted numerically.

\subsection*{Trap fabrication and characteristics}
We use sapphire substrates of size 5\,mm $\times$ 5\,mm $\times$ 2\,mm. During the printing process, a negative-tone acrylate-based photoresist (IP-S, Nanoscribe GmbH \& Co. KG) is placed on top of the substrate. A femtosecond laser of 780\,nm is focused inside the liquid photoresist through a directly immersed 25$\times$ objective (numerical aperture 0.8). The laser focal spot is scanned in 3D to cross-link the photoresist in the designed trap geometry. After 3D printing, the remaining liquid photoresist is washed off by soaking in a developer (propylene glycol methyl ether acetate, Sigma-Aldrich) for four hours and rinsed with isopropanol. After fully drying, the samples are coated with a nominally 1\,\micrometer thick Au-film using an evaporation angle of 30 degrees with respect to the normal of the rotating substrate surface. 

From the observed storage times of a few hours, we deduce that the pressure in the trap center is well below $10^{-10}$\,mbar. Considering that the apparatus was baked for only 4 days at 180$^\circ$C, we conclude that outgassing of the photoresist is sufficiently well suppressed. We also simulate the structural response of the trap electrodes to the electrostatic forces from the applied voltages and find that deformation of the trap electrodes should be well below 0.2\,nm at 200\,V DC voltage. 

\subsection*{Measurements}
Ions are created in the trap by photoionizing neutral $\rm{Ca}$ atoms thermally evaporated from a $\rm{Ca}$ oven. We confine single $^{40}\rm{Ca}^{+}$ ions 130\,\micrometer above the DC electrode plane. The planar motional modes are cooled using a horizontal (parallel to the substrate surface) 397\,nm beam detuned by 20\,MHz to the red from the $4^2 S_{1/2}-4^2 P_{1/2}$ cooling transition while a vertically oriented 866\,nm repumper beam on the $3^2 D_{3/2}-4^2P_{1/2}$ transition provides some partial cooling of the vertical axial mode. 
The planar trap frequencies are measured by sideband spectroscopy on the $\lvert 4^{2} S_{1/2}, 
m_j=-1/2\rangle$ $\leftrightarrow$ $
\lvert 3^{2} D_{5 / 2}, m_j=-5/2\rangle$ 
Zeeman transition using 729\,nm light in a horizontal configuration.

We measure the amplitudes of each RF electrode pair with capacitive dividers, each with a ratio of 3\,pF:100\,pF, and balance them using additional tunable capacitors \cite{an_surface_2018}. DC electrodes are configured to create vertical confinement as well as removing the degeneracy between the two planar motional modes parallel to the substrate.

\end{document}